# Image Inpainting Using Sparsity of the Transform Domain

H. Hosseini*, N.B. Marvasti, *Student Member, IEEE,* F. Marvasti, *Senior Member, IEEE*

*Advanced Communication Research Institute (ACRI)*

*Department of Electrical Engineering, Sharif University of Technology*

hosein.hosseini@gmail.com, neda.marvasti@gmail.com, marvasti@sharif.edu

*Abstract*—In this paper, we propose a new image inpainting method based on the property that much of the image information in the transform domain is sparse. We add a redundancy to the original image by mapping the transform coefficients with small amplitudes to zero and the resultant sparsity pattern is used as the side information in the recovery stage. If the side information is not available, the receiver has to estimate the sparsity pattern. At the end, the recovery is done by consecutive projecting between two spatial and transform sets. Experimental results show that our method works well for both structural and texture images and outperforms other techniques in objective and subjective performance measures.

## I. INTRODUCTION

IMAGE inpainting is the filling-in of missing or undesired portions of an image. Over the past decade, image restoration has become an interesting field of image processing and many techniques have been developed to address this problem.

Image inpainting techniques are categorized into two groups. The PDE-based methods [1]-[3] use the information of the surrounding area of the missing region. These methods are suitable to restore small loss patterns; however they cause smoothing effects in relatively large missing regions. In this case, the exemplar-based techniques [4]-[7] which duplicate the image information from the known region into the missing region at the patch level, results in better restoration. The former group works well for structural images, while the latter is suitable for texture ones.

Also, image features in the transform domain provides an effective tool for image restoration. References [8] and [9] combined the frequency and the spatial domain information to iteratively remove scratches from still images. In [10] the authors propose a technique for image inpainting based on the theory of POCS, exploiting the frequency-spatial representation provided by wavelets.

Here, we use the property that much of the image information in the transform domain is sparse [11]. A sparsity operator maps the transform coefficients with amplitudes, smaller than a pre-specified threshold to zero. The resultant is referred to as the *sparse image* and the location of zeros is the *sparsity pattern.* The sparsity operator adds a redundancy to the original image and the sparsity pattern is used as the *side information* in the recovery stage.

The sparsity information of an image along with the information of the uncorrupted pixels motivates us to apply the theory of projection onto convex sets (POCS) [9] to achieve the final result.

The rest of this paper is organized as follows. Section II provides the preliminaries background for the techniques used in the proposed method. Our algorithm is presented in section III. The experimental results and comparisons are provided in Section IV and section V concludes the paper.

## II. Preliminaries

In this section, we present a brief review of the techniques used in our proposed method.

### A. Time Varying Method (TV):

The TV is an efficient method to reconstruct a signal from its non-uniformly spaced samples. Assume:

$$x_M(t) = x(t).M(t) \tag{1}$$

where $x(t)$, $M(t)$ and $x_M(t)$ are the original image, the loss mask and the corrupted image respectively and the symbol $(.)$ denotes the pixel-wise multiplication. The loss mask is a binary matrix which is zero in the missing pixels. The image is reconstructed as follows [12]:

$$x_r(t) = \frac{x_M^{lp}(t)}{M^{lp}(t)} \tag{2}$$

where $lp$ denotes the low-pass filtering and $x_r(t)$ is the reconstructed image. Fig. 1 shows the block diagram of this method.

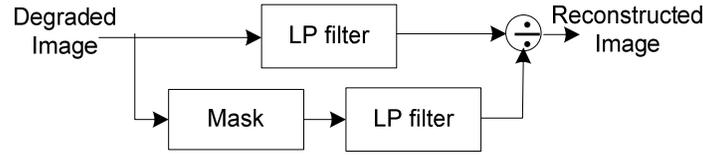

Fig. **1**. The Time-Varying method.

*B. Projection Onto Convex Sets:*

Alternating projections onto convex sets (POCS) is a powerful tool for signal and image restoration [13]. The primary result of POCS is that, consecutive projecting among two or more convex sets with nonempty intersection results in convergence to a point included in the intersection [9]. This is illustrated in Fig. 2.

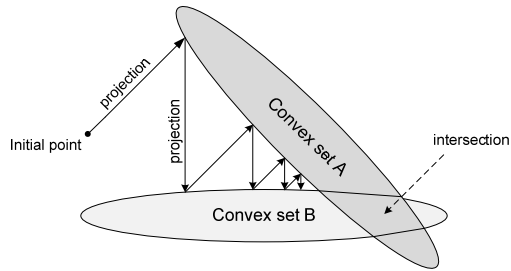

Fig. **2**. Consecutive projecting among two or more convex sets with nonempty intersection results in convergence to a point included in the intersection.

*Definition 1*: A set of signals, $A$ is convex if, for $0 \leq \alpha \leq 1$, the signal:

$$\alpha\, u_1(x) + (1 - \alpha)u_2(x) \in A \tag{3}$$

where $u_1(x), u_2(x) \in A$ [13].

*Definition 2*: Projection operation is denoted by P. The notion

$$w(x) = P_K v(x) \tag{4}$$

means that $w(x)$ is the projection of $v(x)$ onto the convex set $A_K$. If $v(x) \in A$, then $P_K v(x) = v(x)$.

The two convex sets used in our algorithm are described below:

1- *Sparsity in the Transform Domain*:

The set of sparse signals in the transform domain is:

$$A_S = \{u(x)|\ U(\omega) = 0, \quad for\ \omega \in \omega_S\} \tag{5}$$

where $\omega_S$ is the sparsity pattern.

The set $A_s$ is convex because $\alpha\ u_1(x) + (1 - \alpha)\ u_2(x)$ is a sparse signal when both $u_1(x)$ and $u_2(x)$ have the same sparsity pattern. A signal $v(x)$ is projected onto the set $A_S$, as follows:

$$w(x) = P_S\ v(x)$$

$$= T^{-1}.S.T \tag{6}$$

where $T$ is the transform and S is the sparsity operator as defined below:

$$S(V(\omega)) = \begin{cases} 0 & \omega \in \omega_S \\ V(\omega) & \omega \notin \omega_S \end{cases} \tag{7}$$

The sparsity pattern is the set of transform coefficients with amplitudes, smaller than the threshold. Fig. 3 depicts how to obtain the sparsity pattern and apply the sparsity projection ($SP$).

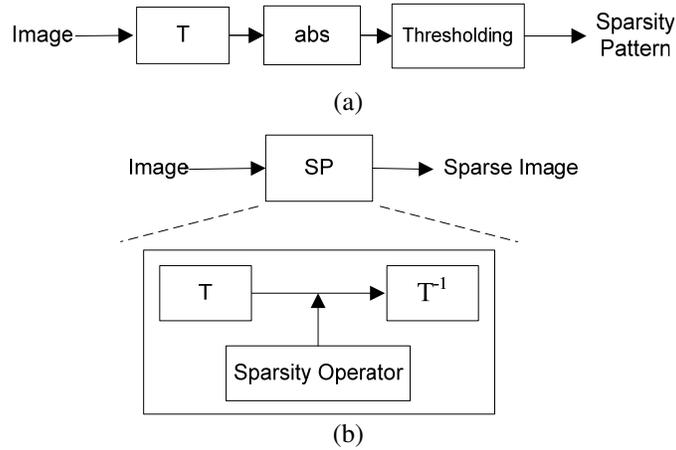

Fig. 3. (a) Obtaining sparsity pattern, (b) Sparsity projection.

2- *Uncorrupted Samples in the Spatial Domain:*

Assume that $x_d$ is the set of uncorrupted samples in the spatial domain and the uncorrupted data are denoted by $I(x)$. The set of signals in the spatial domain:

$$A_D = \{u(x)|\ u(x) = I(x), \quad for\ x \in x_d\} \tag{8}$$

is convex because $\alpha\ u_1(x) + (1 - \alpha)\ u_2(x)$ is an element of $A_D$ when both $u_1(x)$ and $u_2(x)$ belong to $A_D$. A signal $v(x)$ is projected onto the set $A_D$, as follows:

$$w(x) = P_D\ v(x)$$

$$= \begin{cases} I(x) & x \in x_D \\ v(x) & x \notin x_D \end{cases} \quad (9)$$

The constraints that we put on the image and its transform, force them to belong to the corresponding convex sets [9]. The desired result is obtained after several projections.

III. THE PROPOSED METHOD

In our proposed method, two scenarios are considered according to the availability of the image sparsity in the recipient. We describe them below:

*A. Image Inpainting with the Side Information*:

Here, we introduce the image recovery using the side information. The general idea comes from the coding theory. To achieve error correction, the sender has to add some redundancy (i.e., some extra data) to the message, which receiver can use to recover data determined to be erroneous. In our method, the sparsity operator adds the redundancy to the image.

Suppose that the sender transmits the sparse image and shares the sparsity pattern via a reliable channel. In the receiver, the image inpainting (error correction) is done by applying the POCS strategy between two convex sets defined in section II. Fig. 4 illustrates the block diagram of the inpainting stage.

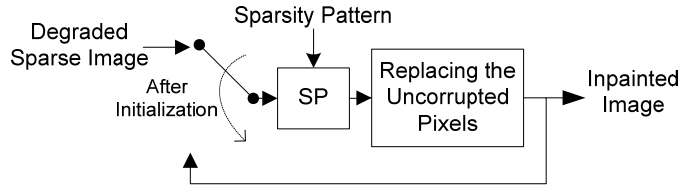

Fig. **4**. The block diagram of the inpainting stage.

The capability of the loss recovery depends on the sparsity percentage of the transmitted image [11]. More sparsity enables the receiver to recover larger losses, but it tends to blur the transmitted image. Hence, the sender should decide on this trade-off, according to the channel parameters.

*B. Image Inpainting without the Side Information*:

In many applications, it is not possible to obtain any information about the original image; for example, the loss may occur during the image acquisition or an object removal may be intended. Thus, in this scenario the receiver has to estimate the side information. For this, the TV method is used to reconstruct the degraded image in order to make an estimation of its sparsity pattern.

Thus, the method in this scenario is as follows:

- The degraded image is reconstructed and the sparsity pattern is estimated,
- The sparsity projection and the loss mask of the degraded image are applied to the reconstructed image to obtain a degraded sparse image,
- The inpainting stage shown in Fig. 4 is applied to the degraded sparse image,
- The uncorrupted pixels of the received image replace the corresponding pixels of the inpainted image.

## IV. SIMULATION RESULTS

In simulations, we use both structural and texture images with a mixture of text and block losses. Several images are used to investigate the performance of the proposed method. In order to quantitatively compare the results, the Peak Signal-to-Noise Ratio (PSNR) between the original and the inpainted images are evaluated.

We exploit the sparsity either in the DCT and the FFT domains. Here we discuss the simulation results for each scenario.

*A- Image Inpainting with the Side Information*:

Table I demonstrates the PSNR value of the inpainted images with respect to the original and the sparse ones for three images *Lena*, *Barbara* and *Baboon* with the missing $16 \times 16$ blocks and with different sparsity percentages. More sparsity results in more similarity to the sparse image and less similarity to the original one, i.e. we can achieve better recovery in the missing region at the expense of lower quality in other parts. However, the transmitted image should have enough sparsity to secure stable recovery in the receiver, e.g. in the FFT domain, 90% sparsity was not enough to recover the corrupted sparse *Lena* image. In this case, sparsity in the DCT domain works better than the FFT. The results are obtained after 500 POCS iterations, except for the cases determined in Table I, which need 1000 iterations. Fig. 5 shows the results for the image *Lena* using 95% sparsity in the DCT domain.

*A- Image Inpainting without the Side Information*:

Fig. 6 shows three images used in the simulations[1]. The LPF operator of the TV method is implemented by progressive convolution of the input with a window defined below:

$$window = \frac{1}{5}\begin{bmatrix} 0 & 1 & 0 \\ 1 & 1 & 1 \\ 0 & 1 & 0 \end{bmatrix} \quad (10)$$

To investigate the performance of the estimator, the sparsity pattern of the recovered image is compared with the original one. For this, we calculate the percentage of undetected sparse coefficients ("miss-detection" terms) and coefficients identified as sparse in error ("false-alarm" terms). The number of miss-detection terms and false-alarm terms are equal, because both the original and the recovered images are assumed to have the same sparsity percentages. For three corrupted images shown in Fig. 6, these values are 0.7%, 1.3% and 1.2% respectively, which indicate that the recovery stage provides a good estimation of the original sparsity pattern.

In this scenario, the sparsity in the FFT domain has better performance. Table II shows the results. We compare our method with the Total-Variation method [3], simultaneous texture and structure inpainting algorithm [4], Criminisi's exemplar-based algorithm [5], Wong's exemplar-based method [6] and inpainting using patch sparsity [7]. The results are obtained after 400 POCS iterations and the corresponding sparsity percentages for three images are 95%, 90% and 85%. Our method, overcomes the smoothing effects of the total-variation method and unwanted objects of the exemplar-based technique. The relative CPU time is of the order of the total-variation method and 0.01 ~ 0.1 of the other techniques.

V. CONCLUSION

In this paper we showed that using side information can help the image restoration. In the proposed method, the sparsity operator adds a redundancy to the original image and the resultant sparsity pattern is used as the side information in the recovery stage. Large missing regions can be recovered at the expense of blurring the transmitted image. Due to the special sparsity pattern of the natural images, the compressed version introduces small overhead to the transmitted data. The sparsity pattern may be included in the header data of the compressed sparse images. It suggests an efficient approach to combat the block loss in the compression schemes. In transform-based compression methods like JPEG, it is recommended to combine the sparsity and compression stages to achieve more efficiency. Also we proposed a method to estimate the sparsity pattern if the side information is not available in the recipient. Our method overcomes the smoothing effects of the PDE-based methods in texture images and unwanted objects of the exemplar-based techniques in structural images.

---

[1] The images and the results of [7] are used.

TABLE I

THE PSNR VALUE OF THE INPAINTED IMAGES WITH RESPECT TO THE ORIGINAL AND THE SPARSE ONES FOR THREE IMAGES WITH THE MISSING 16 × 16 BLOCKS AND WITH DIFFERENT SPARSITY PERCENTAGES. THE RESULTS ARE OBTAINED AFTER 500 POCS ITERATIONS.

|   | FFT | | | DCT | | |
|---|---|---|---|---|---|---|
|   | Sparsity Percentage | PSNR with respect to the original image | PSNR with respect to the sparse image | Sparsity Percentage | PSNR with respect to the original image | PSNR with respect to the sparse image |
| *Lena* | 90% | 25.6 dB | 32.3 dB* | 90% | 33.8 dB | 42.7 dB |
|  | 95% | 30.0 dB | 40.0 dB | 95% | 31.5 dB | 63.5 dB |
| *Barbara* | 90% | 27.6 dB | 40.9 dB | 85% | 30.7 dB | 51.8 dB |
|  | 95% | 25.6 dB | 62.4 dB | 90% | 28.6 dB | 62.7 dB |
| *Baboon* | 90% | 23.2 dB | 33.5 dB* | 75% | 28.8 dB | 47.0 dB |
|  | 95% | 22.2 dB | 61.7 dB | 80% | 27.5 dB | 56.3 dB |

*1000 POCS iterations.

TABLE II

COMPARISON OF THE INPAINTING RESULTS IN PSNR FOR THREE IMAGES OF THE FIG. 6.

|   | *Total-Variation* [3] | *Bert* [4] | *Crim* [5] | *Wong* [6] | *Spar* [7] | *Our Method* |
|---|---|---|---|---|---|---|
| (a) | 19.1 dB | 20.1 dB | 18.4 dB | 20.5 dB | 20.6 dB | **21.2 dB** |
| (b) | 23.4 dB | 22.8 dB | 21.6 dB | 23.3 dB | 23.8 dB | **24.9 dB** |
| (c) | 25.5 dB | 25.2 dB | 21.2 dB | 24.8 dB | 25.6 dB | **26.1 dB** |

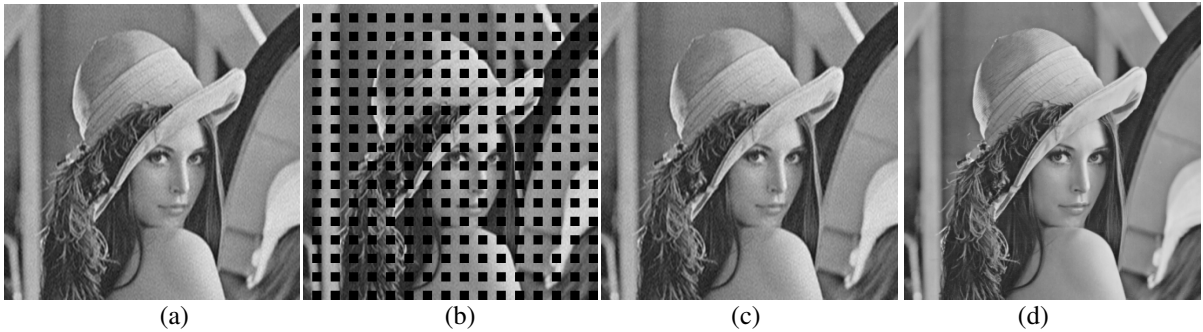

(a) (b) (c) (d)

Fig. **5**. Results for the image Lena using 95% sparsity in the DCT domain, (a) The sparse image, (b) The corrupted sparse image, (c) The inpainted image, and (d) The original image.

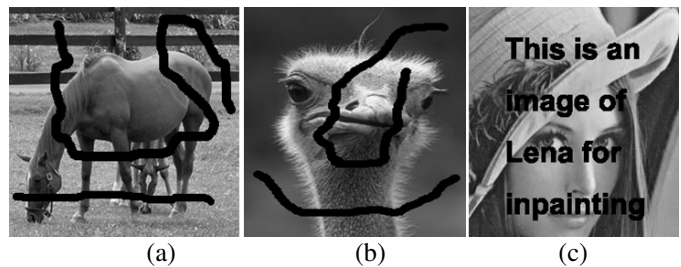

(a) (b) (c)

Fig. **6**. Three corrupted images used in the simulations.